\pdfoutput=1
\PassOptionsToPackage{dvipsnames}{xcolor}
\documentclass{SciPost}

\hypersetup{
    colorlinks,
    linkcolor={red!50!black},
    citecolor={blue!50!black},
    urlcolor={blue!80!black}
}

\usepackage[bitstream-charter]{mathdesign}
\DeclareSymbolFont{usualmathcal}{OMS}{cmsy}{m}{n}
\DeclareSymbolFontAlphabet{\mathcal}{usualmathcal}

\fancypagestyle{SPstyle}{
\fancyhf{}
\lhead{\colorbox{scipostblue}{\bf \color{white} ~SciPost Physics }}
\rhead{{\bf \color{scipostdeepblue} ~Submission }}

\fancyfoot[C]{\textbf{\thepage}}
}

\newcommand{\figref}[2]{\hyperref[#1]{\autoref*{#1}(#2)}}

\usepackage{bm}
\usepackage{subcaption}
\usepackage{placeins}
\usepackage{tikz}
\definecolor{Jhcol}{RGB}{0,25,162}    
\definecolor{Jtcol}{RGB}{42,157,143}  
\definecolor{Jdcol}{RGB}{232,160,44}  
\usepackage[normalem]{ulem}
\usepackage{cleveref}

\graphicspath{{../images/}}
\usepackage[color=violet!30]{todonotes}

\renewcommand{\Re}{\operatorname{Re}}

\newcommand{\Tr}{\mathrm{Tr}}

\begin{document}

\pagestyle{SPstyle}

\begin{center}{\Large \textbf{\color{scipostdeepblue}{
Sign-optimized Quantum Monte Carlo
}}}\end{center}

\begin{center}\textbf{
Julius S. Herz\textsuperscript{1$\star$},
Robin Schäfer\textsuperscript{2},
Mat\'ias G. Gonzalez\textsuperscript{1}, and
David J. Luitz\textsuperscript{1}
}\end{center}

\begin{center}
{\bf 1} Institute of Physics, University of Bonn, Meckenheimer Allee 166a, 53115 
Bonn, Germany
\\
{\bf 2} Department of Physics, Harvard University, Cambridge, MA 02138, USA
\\[\baselineskip]
$\star$ \href{mailto:jherz@uni-bonn.de}{\small jherz@uni-bonn.de}
\end{center}

\section*{\color{scipostdeepblue}{Abstract}}
\textbf{\boldmath{%
The sign problem breaks the polynomial scaling of Quantum Monte Carlo methods.
We alleviate it by rotation of the local basis of the Hilbert space, such
that the phase of off-diagonal matrix elements of the rotated Hamiltonian is 
minimized. These minima coincide with optima of the average sign.
We benchmark our method for frustrated Heisenberg antiferromagnets in one 
dimension and on the two-dimensional maple-leaf lattice. Our approach reveals 
more efficient bases with improved sign in a large part of the parameter space 
for all models considered, enabling us to reach lower temperatures, on par with 
state-of-the-art numerical linked cluster expansions which we directly compare 
to.}}

\vspace{\baselineskip}

\noindent\textcolor{white!90!black}{%
\fbox{\parbox{0.975\linewidth}{%
\textcolor{white!40!black}{\begin{tabular}{lr}%
  \begin{minipage}{0.6\textwidth}%
    {\small Copyright attribution to authors. \newline
    This work is a submission to SciPost Physics. \newline
    License information to appear upon publication. \newline
    Publication information to appear upon publication.}
  \end{minipage} & \begin{minipage}{0.4\textwidth}
    {\small Received Date \newline Accepted Date \newline Published Date}%
  \end{minipage}
\end{tabular}}
}}
}


\vspace{10pt}
\noindent\rule{\textwidth}{1pt}
\tableofcontents
\noindent\rule{\textwidth}{1pt}
\vspace{10pt}

\section{Introduction}
\label{sec:introduction}

There is no universal, unbiased, and efficiently scalable method to solve 
generic quantum many-body problems. While some problems can be tackled in 
polynomial time by quantum Monte Carlo methods, this is not true in general. In 
particular, exotic phases of matter often emerge from competing interactions, 
which can introduce frustration and hence render stochastic approaches 
ineffective due to the infamous \emph{sign problem}. For example, in spin 
systems, geometric frustration can induce a wide range of unconventional 
emergent collective behaviors~\cite{Diep2005, Balents2010, Savary2016}. Such 
systems exhibit emergent properties including pinch-point singularities 
associated with emergent gauge structures~\cite{Fennell2009, Henley2010, 
LozanoGomez2025}, photon-like excitations~\cite{Hermele2004, Gao2025}, magnetic 
monopoles~\cite{Castelnovo2008, Morris2009}, and fractionalized spin excitations 
such as spinons~\cite{Broholm2020}.

Interestingly, in such cases, also other exact methods typically cannot reach 
the scaling limit.
For this reason, the development of a complementary set of techniques is an 
ongoing effort.
On one end of the spectrum, exact diagonalization and the density-matrix 
renormalization group (DMRG) algorithm provide powerful exact or quasi-exact 
tools~\cite{White1992,Schollwock2005}, particularly in one dimension and zero 
temperature. DMRG has also achieved notable success in certain two- and 
three-dimensional systems~\cite{Schollwock2011,Schafer2021}, but it struggles 
when the entanglement entropy is large. Other approaches rely on physical insight, 
such as variational Monte Carlo methods~\cite{Becca2017}, or series expansions 
constructed around solvable limits~\cite{Oitmaa2006}. Other methods are exact in 
the infinite-temperature limit, such as high-temperature series 
expansions~\cite{Oitmaa2006, Bernu2020, Gonzalez2022}, the pseudo-Majorana 
functional renormalization group method~\cite{Niggemann2021, Niggemann2022, 
Schneider2024, Bippus2025}, and numerical linked-cluster expansions 
(NLCE)~\cite{rigol_numerical_2006,rigol_numerical_2007I,rigol_numerical_2007II,Schafer2020,schaefer_magnetic_2022}, among others. However, these methods tend to 
struggle with convergence at low and very low temperatures as the correlation length
grows.

It is hence highly desirable to restore the power of quantum Monte Carlo methods
in cases suffering from the sign problem. The sign problem is not an
intrinsic property of a system, but depends on the representation in which the
simulation is carried out. A generic solution is nevertheless out of reach, as
removing the sign problem for an arbitrary Hamiltonian is an NP-hard
task~\cite{Troyer2005}. Substantial effort has therefore been devoted to
alleviating the sign problem in various systems~\cite{Nakamura1998,
Shinaoka2015, Hangleiter2020, Pan2024, Murota2025,
Wynen2021, Rodekamp2022, Chandrasekharan1999, Li2015, Alexandru2022}, most
commonly by identifying, guided by physical intuition, a favorable simulation
basis for the model at hand. There are several
cases in which, surprisingly, the sign can be cured or recovered at low 
temperatures by separating the system into clusters and using the eigenbasis of 
the cluster to perform calculations~\cite{Alet2016,Honecker2016,Wessel2017, DEmidio2020, Honecker2022, 
Weber2022a, Weber2022b}. 

In this article, we propose an improved method for easing the QMC sign problem 
through a basis optimization procedure guided by a generalization of the 
non-stoquasticity of the bond Hamiltonian. The non-stoquasticity measures the 
size of the negative elements in the bond Hamiltonian, and was previously used 
as a cost function in Ref.~\cite{Hangleiter2020}. Here, we extend this approach
to the full space of local bases, reached by local unitary rotations.
It is only through this generalization that we recover the known limit for bipartite systems, where a Marshall rotation renders the Hamiltonian stoquastic (sign-problem free) \cite{Marshall1955}.

We establish non-stoquasticity as a useful cost function in quasi-one- and two-dimensional frustrated models:
\emph{(i)} it is noise-free, because it can be computed without a full QMC 
calculation, \emph{(ii)} its local minima appear to coincide with the maxima of 
the sign in the cases we investigated, \emph{(iii)} its gradient seems to be 
aligned with the gradient of the sign during optimization.

\begin{figure}[t]
    \centering
    \begin{minipage}[t]{0.49\textwidth}
        {\raggedright\hspace{0.5em}\textbf{a)}\par}
        \centering
        \vspace{1em}
        \resizebox{0.9\linewidth}{!}{\input{figs/fig_mapleaf.tikz}}
    \end{minipage}\hfill
    \begin{minipage}[t]{0.49\textwidth}
        {\raggedright\hspace{1.5em}\textbf{b)}\par}
        \centering
        \includegraphics[width=\linewidth]{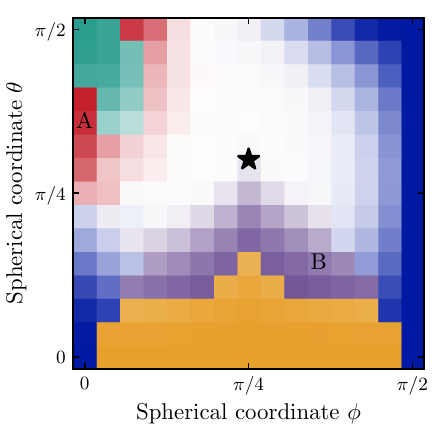}
    \end{minipage}
    \caption{(Left) Maple-leaf lattice with its three symmetry-inequivalent
    couplings: $J_d$ (amber) along dimers, $J_t$ (teal) along triangles, and $J_h$
    (blue) along hexagons. (Right) Computational phase diagram of the MLL for
    $T=0.5J$: the color indicates the basis with the largest average QMC sign
(computational/dimer/trimer eigenbasis in blue/amber/teal; optimized 2-spin/3-spin
clustering in purple/red) on an $N=54$ site lattice. The transparency indicates the average sign. The isotropic point
	$J_h=J_t=J_d$ is marked by the black star.}
    \label{fig:1}
\end{figure}

We apply our
optimization scheme to the $S=\frac 1 2$ Heisenberg antiferromagnet on the two-dimensional frustrated maple-leaf lattice (MLL) and find that it
significantly mitigates the sign problem across a large portion of the phase diagram, as shown in \figref{fig:1}{b}.
The MLL contains three symmetry-inequivalent bonds: hexagonal (blue), triangular (teal), and dimer (amber).
We parametrize the strength of the corresponding $SU(2)$-symmetric couplings, $(J_h, J_t, J_d)$, using the spherical coordinates $(\phi,\theta)$ introduced in Ref.~\cite{Ebert2026}.
The transparency scale in \figref{fig:1}{b} shows for each parameter point the best obtained average sign across five classes of bases.
The color indicates in which class of bases the optimal sign is found.
We minimize our generalized cost function by optimal choice of local unitary rotations acting on two-site (purple) and three-site (red) clusters. 
The corresponding two- and three-site cluster bases include, as special cases, the eigenbases of the dimer (amber) and trimer (teal), respectively.
Additionally, regions where the computational basis yields the highest average sign are shown in blue.
Fully transparent (white) regions correspond to an average sign close to zero, where the sign problem is maximal.
We identify large regions where the optimized bases are neither computational nor local eigenbases.

The remainder of the article is organized as follows: In \autoref{sec:methods}, we introduce the stochastic series expansion formulation of QMC and the basis-optimization procedure. \autoref{sec:trilad} contains the applications to ladder systems and in \autoref{sec:maplat} we study the phase diagram of the maple-leaf lattice. Finally, \autoref{sec:conclusion} summarizes our results and discusses their implications.

\section{Quantum Monte Carlo}
\label{sec:methods}

Here we introduce the basis-independent formulation of the stochastic series expansion (SSE) QMC and present the details of our optimization method. 

\subsection{Stochastic series expansion and the sign problem}

The SSE method~\cite{Sandvik2019} is based on the expansion around the infinite-temperature limit of the partition function $Z = \Tr[\exp(-\beta H)]$, where $H$ is the Hamiltonian and $\beta$ the inverse 
temperature. Writing out the trace in an arbitrary basis $\{|\alpha\rangle\}$, we 
obtain
\begin{equation}
    \label{eq:z}
    Z = \sum_{n=0}^{\infty} \sum_{\alpha} \langle \alpha | \frac{(-\beta)^n}{n!} H^n | \alpha \rangle.
\end{equation}
Two-local Hamiltonians can be decomposed into a sum over the bond operators $h_b$ of the lattice, as
\begin{equation} 
H = -\sum_b h_b, 
\label{eq:h_bond_decomp}
\end{equation}
where $b$ denotes a bond connecting lattice sites $i$ and $j$. Note that we have added an additional minus sign in \autoref{eq:h_bond_decomp} in order to absorb the $(-1)^n$ factor in \autoref{eq:z}. For the Heisenberg model we will study here, we have $h_b = -J\, 
\mathbf{S}_i \cdot \mathbf{S}_j + \frac{J}{4} I$. The physically irrelevant constant makes the diagonal elements of $h_b$ non-negative.
Generalization to other models is straightforward. We further split $h_b$ into the diagonal ($x=0$) and off-diagonal $(x=1)$ terms $h_b = \sum_{x\in\{0,1\}} h_{b,x}$. Plugging this into the partition function yields
\begin{equation}
Z = \sum_{\alpha,n} \sum_{b_1,\dots,b_n} \sum_{x_1,\dots,x_n} \frac{\beta^n}{n!} 
\langle \alpha |h_{b_1,x_1} \cdots h_{b_n,x_n}| \alpha \rangle,
\label{eq:z_expanded}
\end{equation}
where the sum can now be viewed as a sum over all \emph{configurations} $C_n=\{\alpha, (b_1,x_1,\cdots,b_n,x_n) \}$, which are uniquely determined by the trace state $\alpha$ and the ordered sequence of bond operators, called \emph{operator string}, whose length $n$ defines the \emph{expansion order}. Thus, 
\begin{equation} 
Z = \sum_{C_n} W(C_n),
\end{equation}
where each configuration defines a weight $W(C_n)$ which can be interpreted as the statistical weight of a 
given Monte Carlo configuration, allowing to perform an importance sampling over 
different configurations $C_n$ to construct a Markov chain. Within this 
framework, the sampling algorithm consists of two types of updates: diagonal 
updates, which remove and insert diagonal operators $h_{b,0}$; and off-diagonal updates, 
which change the state $|\alpha\rangle$ and can transform the types of 
operators. We employ the general sampling algorithm presented in 
Refs.~\cite{Weber2022a, Alet2005}, which is not restricted to the single-spin $S_i^z$
eigenbasis. Throughout, we refer to this $S_i^z$ eigenbasis as the \emph{computational} basis.

The negative-sign problem arises when the weights $W(C_n)$ are not positive definite, causing the stochastic interpretation that SSE relies on to break down. Regardless, one 
may still calculate expectation values of observables $O$ via reweighting as
\begin{equation}
\langle O \rangle = \frac{\langle O \, s \rangle_{|W|}}{\langle s \rangle_{|W|}} ,
\label{eq:reweighting}
\end{equation}
where $s$ is the sign or phase in the complex case of a given weight, $s(C_n)=\frac{W(C_n)}{|W(C_n)|}$ and $\langle \cdot \rangle_{|W|}$ denotes the Monte Carlo expectation value, i.e.\ the average taken along the Markov chain whose configurations $C_n$ are sampled with probability proportional to $|W(C_n)|$~\cite{Loh1990,Pan2024}. The distribution $|W(C_n)|$ and the associated partition function $Z^\prime$ correspond to the so-called unsigned Hamiltonian, obtained by making all matrix elements of $h_b$ in \autoref{eq:h_bond_decomp} non-negative. Since $W(C_n)$ is basis dependent, so is the unsigned Hamiltonian. Given this, the average sign or phase $\langle s \rangle$ can be calculated as~\cite{Troyer2005,Pan2024, Hatano1992}
\begin{equation}
\langle s \rangle =\frac{Z}{Z^\prime}= e^{-\beta N \Delta f},
\label{eq:average_sign_formula}
\end{equation} 
where $\Delta f$ is the free energy density difference between the signed and unsigned systems. This leads to exponentially large statistical fluctuations when $\Delta f$ is non-zero, making simulations computationally infeasible at large $N$ and low temperatures.

The QMC-SSE formulation for Heisenberg antiferromagnets on non-bipartite 
lattices is bound to suffer from the sign problem (in the computational basis) 
due to the possibility to have operator strings with odd numbers of off-diagonal 
operators related to the perpendicular component $S_i^+ 
S_j^-$~\cite{Henelius2000,Troyer2005} through winding around loops of odd length 
in the lattice. However, since the QMC sign is not a physical quantity, the 
severity of the sign problem depends on the basis in which the simulation is 
performed~\cite{Weber2022a, DEmidio2020, Hangleiter2020, Pan2024,Hatano1992}. 
For example, for the fully frustrated ladder the sign problem can be completely 
cured by working in the two-site rung eigenbasis~\cite{Wessel2017} (see \autoref{sec:ffl}). 

\subsection{Sign optimization}
\label{ssec:sign}

Here, we aim to develop a general scheme that allows alleviation of the negative-sign problem in SSE-QMC by performing local basis rotations on small clusters. But before this, let us first summarize the fundamental works upon which we will build our method. On one hand, there are several studies showing that the sign problem can be alleviated or cured by performing simulations in the eigenbasis of 
a cluster (or supersite) into which the lattice is decomposed~\cite{Wessel2017, DEmidio2020, Honecker2022, Weber2022a, Weber2022b}. This approach is typically effective when the system is close to the limit of isolated clusters. On the other hand, other studies have proposed to perform basis rotations guided by different optimization quantities to find an optimal basis for the QMC simulations~\cite{Levy2021,Hangleiter2020} (the average sign itself is too costly and makes the whole process inefficient). In particular, Ref.~\cite{Hangleiter2020} proposed to use the
computationally-inexpensive and system-size-independent \emph{non-stoquasticity} $\nu$ of the (bond) Hamiltonian, defined as
\begin{equation}
\begin{aligned}
    \nu(O) = \sum_{i \neq j} \frac{1}{2} \left( \left| [O^T h_b O]_{ij} 
    \right| - [O^T h_b O]_{ij} \right), \quad \text{with} \quad O \in 
    \mathrm{O}(d)
\end{aligned}
\label{eq:non-stoq}
\end{equation}
where $[O^T h_b O]_{ij}$ denotes the $(i,j)$ matrix element of the bond 
Hamiltonian $h_b$ in the basis defined by the orthogonal rotation $O\in\mathrm{O}(d)$ relative to the computational basis. The heuristic reason for this cost function is that bringing the Hamiltonian closer to a stoquastic representation (smaller $\nu$) implies smaller negative off-diagonal matrix elements in the bond Hamiltonian and therefore lower chances that operator strings with negative weight appear.

However, stoquasticity ($\nu= 0$) is only a sufficient but not a necessary condition for the absence of the sign problem. For example, the bond Hamiltonian of the Heisenberg antiferromagnet on the square lattice has only negative off-diagonal elements, which would imply a large value of $\nu$. Regardless, the model is sign-problem free due to the lattice being bipartite: the number of bond operators acting off-diagonally is always even and thus all configuration weights remain positive definite~\cite{Sandvik2019}. In fact, the bond Hamiltonians of Heisenberg antiferromagnets on bipartite lattices can always be rendered stoquastic by a unitary Marshall sign transformation on one sublattice~\cite{Marshall1955}.
More generally, the problem of deciding whether a Hamiltonian can be rendered stoquastic by local basis transformations has 
recently been studied systematically in Ref.~\cite{Klassen2019}.

We extend the orthogonal basis rotation to unitary basis rotations, as it is more expressive. Because the elements of the bond Hamiltonian can be complex, we introduce the \emph{generalized non-stoquasticity},
\begin{equation}   
    \mathcal{S}(U) = \sum_{i \neq j} \frac{1}{2} \left( \left| [U^\dagger h_b U]_{ij} \right| - \Re \left([U^\dagger h_b U]_{ij}\right) \right), \quad \text{with} \quad U \in \mathrm{U}(d).
    \label{eq:cost}
\end{equation}
This quantity does not penalize real positive entries, while penalizing real 
negative entries the most and complex entries according to their phase. The cost 
function is non-negative and vanishes if and only if all off-diagonal matrix 
elements of the bond Hamiltonian are real and non-negative, i.e., if the bond 
Hamiltonian is stoquastic. 
To perform local (unitary) basis rotations, we decompose the lattice into subclusters. For spin $S=1/2$ Heisenberg models, a cluster of $n$ spins yields a $2^n$ dimensional \emph{supersite} on an effective lattice. The effective bond operators are constructed as sums of the underlying inter-cluster interactions originating from the original model. Then, we parametrize the unitary transformations such that the optimization can then be done over the corresponding real-valued parameters. For two-site clustering, for example, we parametrize the $\mathrm{U}(4)$ space by a product of (complex) Givens rotations, which are unitaries that act non-trivially only on a two-dimensional subspace~\cite{Clements2017, Reck1994}. The unitary $U(\bm{\theta})\in \mathrm{U}(4)$ is then represented as a finite product of such elementary rotations, with the phases collected in a real vector $\bm{\theta}\in \mathbb{R}^{16}$ (see \autoref{sec:appendix_givens} for more details). For larger supersites, this parametrization becomes cumbersome. Thus, for three-site clusters ($d=8$), we resort to an exponential-map parametrization,
\begin{equation}
    U(\bm{x}) = e^{iH(\bm{x})}, \qquad H(\bm{x}) = H(\bm{x})^\dagger \in \mathbb{C}^{8\times 8},
\end{equation}
where the Hermitian generator $H(\bm{x})$ is constructed from a set of real parameters $\bm{x}\in \mathbb{R}^{64}$. This guarantees $U(\bm{x})\in \mathrm{U}(8)$ exactly, while allowing us to optimize over unconstrained real variables.

To navigate the optimization landscape, we employ a two-stage optimization strategy. In the first stage, we use the adaptive first-order method \emph{Adam} starting from a randomly chosen initial basis rotation~\cite{Kingma2017}. We then run Adam for a fixed number of iterations using a cosine-annealing learning-rate scheduler, which gradually decreases the learning rate during the optimization. This yields an intermediate solution $U^{(\mathrm{Adam})}$. In the second stage, we refine this solution by applying the quasi-Newton limited-memory Broyden–Fletcher–Goldfarb–Shanno algorithm (LBFGS)~\cite{Liu1989}. Starting from $U^{(\mathrm{Adam})}$, LBFGS uses curvature information accumulated from gradient differences to resolve the minimum more accurately than a pure first-order method. The final optimizer is thus obtained as
\begin{equation}
    U^\star = \mathrm{LBFGS}\bigl[\mathrm{Adam}(U_{0})\bigr],
\end{equation}
where $U_{0}$ denotes a random initialization sampled from the Haar 
distribution~\cite{Mezzadri2007}. As an example, we run Adam for around $4000$
iterations, and the subsequent LBFGS stage is completed once the relative
improvement of the cost function $\mathcal{S}$ falls below a predefined threshold
of $10^{-9}$. By repeating the two-stage optimization for multiple random
initializations, we are able to find many different local minima and compare 
them, as we will show below. To complete the method, whenever the effective superlattice is bipartite we
allow for two independent basis rotations, one on each sublattice. The use of
two distinct rotations rather than a single uniform one is motivated by the
Marshall sign transformation: the stoquastic representation of a bipartite
Hamiltonian is reached by a staggered, sublattice-dependent unitary rotation. With this
additional freedom, an optimization of a sign-problem-free model started from a
random basis rotation consistently converges to a stoquastic ($\mathcal{S}=0$)
representation of the Hamiltonian.

\section{Application to ladder systems}
\label{sec:trilad}

We now apply our optimization method to two different two-leg ladder systems, the fully frustrated and the triangular ladders shown in the insets of \autoref{fig:opt_series}. For the former, a sign-problem-free basis is known to exist, and we show that our optimization method reliably recovers it while reducing the non-stoquasticity exactly to zero. For the latter, no local basis rotation is known to cure the sign problem. Here, our method identifies a basis with significantly improved sign performance compared to the computational and cluster eigenbases.

\subsection{Fully frustrated ladder}
\label{sec:ffl}

\begin{figure}[t]
    \centering
    \includegraphics[width=\textwidth]{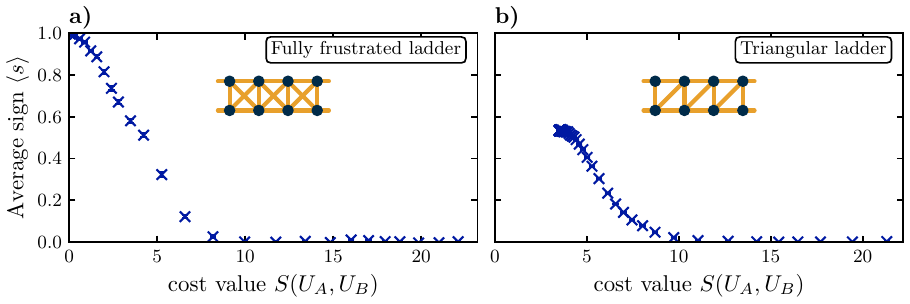}
	\caption{(Left) average QMC sign $\langle s\rangle$ as a function of the 
generalized non-stoquasticity $S(U_A, U_B)$ along the optimization trajectory 
for the fully-frustrated ladder (shown in the inset) of $N=24$ sites at $T=J$. 
(Right) the same for the triangular ladder.}
    \label{fig:opt_series}
\end{figure}

We apply the optimization method to the two-leg ladder Heisenberg antiferromagnet at the critical point between the rung-singlet phase and spin-1 Haldane phase (also 
called fully-frustrated ladder). This is an 
interesting limit because the QMC calculations can be rendered sign-free by 
using the eigenbasis of the vertical rungs for the simulations~\cite{Alet2016, 
Honecker2016, Wessel2017}. This happens because the leg 
and
diagonal couplings have the same strength and the Hamiltonian can be
rewritten solely in terms of the total spin $\vec{T}_i = \vec{S}_i^1 +
\vec{S}_i^2$ of each rung, with the spin-difference contributions $\vec{D}_i =
\vec{S}_i^1 - \vec{S}_i^2$ cancelling exactly~\cite{Wessel2017}. As a
consequence, the total rung spin $\vec{T}_i^2$ becomes a local conserved
quantity on every rung, taking the values $T_i=0$ (rung singlet) or $T_i=1$
(rung triplet). The singlets are inert, whereas the triplets act as effective
$S=1$ moments that interact through a Heisenberg coupling $\vec{T}_i \cdot
\vec{T}_{i+1}$ along the chain of rungs, so that the model maps onto an
effective spin-1 Heisenberg chain. Since this chain is bipartite, and hence unfrustrated, QMC simulations performed in the
rung eigenbasis are free of the sign problem. We consider a two-site cluster basis defined on the vertical rungs
and assume an even number of rungs with periodic boundary conditions, such that
we can use alternating $\mathrm{U}(4)$ rotations $U_A$ and $U_B$ on the
resulting bipartite chain superlattice. The resulting absolute value of the 
(complex-valued) average QMC sign, $\langle s\rangle$, is shown in \figref{fig:opt_series}{a} as a function of the non-stoquasticity 
$\mathcal{S}$ (only a fraction of the 
optimization steps are shown, for clarity). There, we can see that the optimization method produces a
monotonic decrease of the cost function accompanied by an increasing sign average $\langle s\rangle$. At the beginning of the optimization, the average sign remains near zero and only starts to increase
sharply once the cost function falls below a certain threshold (in \autoref{fig:opt_series} at $\mathcal{S}\approx 10$). This shows that the cost function provides a suitable and, more 
importantly, less expensive measure of the average sign.

In contrast to Ref.~\cite{Hangleiter2020}, which optimized over orthogonal (real) basis rotations acting uniformly on all clusters and could thereby not bring the non-stoquasticity down to zero, i.e.\ did not find a fully stoquastic Hamiltonian, we allow for both general complex-valued unitary
rotations and a sublattice dependence through the alternating rotations $U_A$ and
$U_B$. This additional freedom enables us to reduce the cost function exactly to zero, reaching the stoquastic point $\mathcal{S}=0$ as expected for this sign-problem-free model. We also observe that, for other sign-problem-free models (e.g., ferromagnetic systems), our optimization method also works in the sense that it reliably finds the stoquastic representation of the Hamiltonian when initialized randomly.

\subsection{Triangular ladder}

We now apply the optimization scheme to the antiferromagnetic Heisenberg chain with nearest and next-nearest neighbors. This system can also be seen as a zig-zag ladder (see \figref{fig:opt_series}{b}). Here we focus on the case where all exchange couplings are the same, also known as the two-leg triangular ladder. In this limit, there is no known local basis rotation that cures the sign problem. The optimization is performed again on $U_A \times U_B \in \mathrm{U}(4)\times \mathrm{U}(4)$ acting on the super-sites of two spin clusters along the diagonal rungs. We initialized the optimization with 100 different random starting points distributed according to the Haar distribution.

We show in \figref{fig:opt_series}{b} one optimization run, where we observe a qualitatively similar behavior to the previously shown fully frustrated ladder. However, in this case $\mathcal{S}$ does not reach the stoquastic limit $\mathcal{S}=0$. Consequently, $\langle s\rangle$ is not equal to one. It is important to note that, whereas the non-stoquasticity is temperature-independent, the value of the sign $\langle s \rangle$ depends on the temperature and system size, when the sign problem is not completely cured [see \autoref{eq:average_sign_formula}]. Even though each initialization produces a different basis (the Frobenius norm of the difference between any two resulting $\mathrm{U}(4)$ matrices is large), they can be grouped into a handful of families according to their sign performance and non-stoquasticity. It is important to note that each family of bases is not equally populated, implying that some families are more frequently found than others; and the frequency is unrelated to the performance (for more details see \autoref{sec:appendix_histogram}).

\begin{figure}[t]
    \centering
    \includegraphics[width=\linewidth]{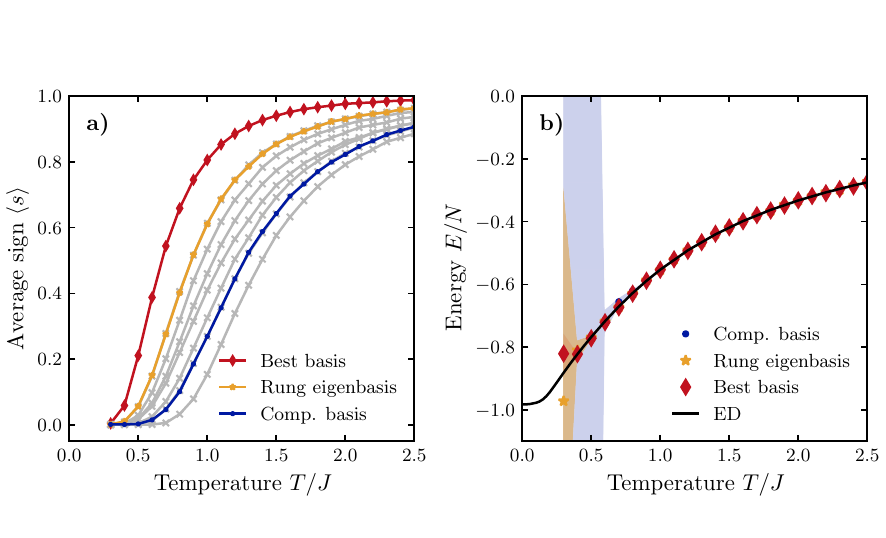}
	\caption{(Left) Average sign against temperature for the triangular 
        ladder system simulated in bases resulting from the optimization with 
    randomly distributed initial conditions. For each class of bases we plot one 
sign--temperature curve. For comparison, the sign-temperature curves for the 
computational basis and the rung eigenbasis are also plotted. (Right) Energy
obtained from QMC simulations in the same bases. The black line 
represents the ED result for an $N=12$ lattice. The simulations are 
performed on an $N=24$ triangular ladder.}
        \label{fig:multiple_triLadder}
\end{figure}

We show in \figref{fig:multiple_triLadder}{a} the temperature dependence of $\langle s \rangle$ for a representative of each family of bases, where the best-performing basis is shown in red diamonds and lines. For comparison, we also add in amber and blue the results for the rung eigenbasis and the computational basis, respectively, which are in agreement with two of the families found. Notably, the best basis we found outperforms the rung eigenbasis and the computational basis, evidencing that our method is capable of finding non-trivial bases which are more convenient for the simulations. We compare in \figref{fig:multiple_triLadder}{b} the resulting energies as a function of the temperature for the best basis we found, the computational basis, and the two-site eigenbasis. For comparison, we show exact diagonalization results in black. As expected, we observe that the best basis produces smaller error bars at lower temperatures for the same computational effort due to the improved sampling efficiency.

Altogether, these results provide a first insight into the optimization landscape. We see that the non-stoquasticity as a cost function exhibits many local minima in the unitary space. It is important to note that the non-stoquasticity of each basis is not directly proportional to its sign performance. For example, the best basis has a higher non-stoquasticity than the supersite eigenbasis, where the average sign drops to zero faster. Thus, the cost function we defined in \autoref{eq:cost} is not a global measure of the average QMC sign and the only global statement we can reliably make is that 
\begin{equation}
    \mathcal{S} (U_A, U_B) = 0 \quad \Rightarrow \quad \text{sign-free QMC simulation}.
\end{equation}

Furthermore, when calculating $\langle s \rangle$ for different system sizes, we do not observe any crossing of the sign-temperature curves of distinct bases for a fixed system size. This suggests that the statement that one basis performs better than another is independent of both temperature and system size. Finally, we point out that we could not identify any particular structure in the optimized unitaries \(U_A, U_B\). They are neither block-diagonal nor equivalent (up to a global phase) to orthogonal matrices. To verify the latter, we checked that the condition $UU^T = e^{i\phi} I$, which would imply that \(U\) is proportional to an element of \(\mathrm{O}(4)\), is not fulfilled. Therefore, the optimal basis cannot be obtained by restricting to orthogonal transformations, indicating that genuinely complex unitary degrees of freedom offer a considerable improvement.

\subsection{Further characterization of the optimization method}
\label{sec:grads}

As we mentioned above, the average sign is too expensive of a cost function and this is why one usually resorts to auxiliary cost functions. Here we will show that the non-stoquasticity is indeed a suitable cost function, as it represents faithfully the local minima in the average sign landscape as well as providing a smoother landscape to guide the optimization process. To illustrate this, we calculate the gradient of the sign using finite differences as
\begin{equation}
\frac{\partial s(\bm{\theta})}{\partial \theta_i}
\approx
\frac{s(\bm{\theta} + \epsilon\,\mathbf{e}_i) - s(\bm{\theta}- \epsilon\,\mathbf{e}_i)}{2\epsilon},
\qquad i = 1,\dots,32,
\end{equation}
where \(s(\bm{\theta})\) denotes the average QMC sign in the basis defined by \((U_A \otimes U_B)(\bm{\theta})\), with \(\bm{\theta}\in \mathbb{R}^{32}\) being the 32 components (which act as \emph{directions}) of the Givens rotation parametrization. Similarly, we can also define the gradient of the non-stoquasticity $\mathcal{S}$ in an equivalent way, with the exception that this does not require any QMC simulation.

\begin{figure}[t]
    \centering
    \includegraphics[width=\textwidth]{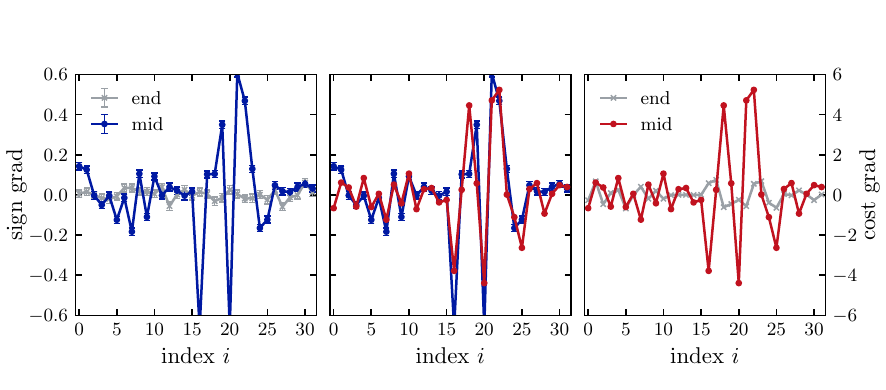}
	\caption{(Left) Sign gradient at the middle and end points of the optimization procedure. (Right) Non-stoquasticity gradient at the middle and end points of the optimization procedure. (Middle) Comparison between the two gradients at the middle point of the optimization (data taken from the left and right panels). Every point $i$ represents a component of the gradient in the 32-dimensional space defined by the parametrization (see main text). We choose $\epsilon = 0.075$. The simulations are performed on an $N=20$ triangular ladder at $T=J$.}
    \label{fig:grad_compa}
\end{figure}

We show the results in \autoref{fig:grad_compa} for two points along the 
optimization process: the end point where the sign is maximal and the 
non-stoquasticity is minimal (in gray color), and a point in the middle where 
the sign is about half of the maximum (blue and red colors). The results for
the sign and non-stoquasticity gradients are displayed on the left and right 
panels, respectively. The gradients at the optimization endpoint are nearly zero 
in both cases. This indicates that local minima of the cost function correspond 
to local maxima of the sign. However, it is relevant to note that there 
\emph{could} be local maxima of the sign that do not correspond to local minima 
of the cost function. On the other hand, at the midpoint, both gradients are 
clearly nonzero and pointing roughly in the same direction, as can be seen in the 
middle panel of \autoref{fig:grad_compa}. This provides numerical evidence that 
the local non‑stoquasticity is a suitable and cheap cost function for the 
optimization process, in the sense that optimizing it simultaneously optimizes 
the sign $\langle s\rangle $. 

Another notable aspect of the non-stoquasticity $\mathcal{S}$ is already visible in \autoref{fig:opt_series}. Starting from a random unitary, the early steps of the optimization process show an improvement on the non-stoquasticity without any apparent change on the average sign. This indicates that $\mathcal{S}$ is able to \textit{see} something in the landscape which $\langle s\rangle$ is still not able to grasp.  In other words, the local minimum of the cost function is broader than the local maximum of the average sign. We confirmed this scenario by calculating the sign $s[\bm{\theta}(\delta)]$ and the cost $\mathcal{S}[\bm{\theta}(\delta)]$ along a straight line $\bm{\theta}(\delta)\in \mathbb{R}^{32}$ through parameter space, originating at $\bm{\theta}(\delta=0)$ in a cost minimum (see \autoref{ch:landscape} for more details).

\section{Application to the maple-leaf lattice}
\label{sec:maplat}

While the previous sections focused on quasi-one-dimensional systems, we now move to a 
challenging frustrated two-dimensional system: The Heisenberg model on the maple 
leaf lattice. We have seen that our basis optimization method can return bases that perform 
better (sign-wise) than the computational basis or the supersite eigenbasis. In 
some cases, the optimization procedure also finds bases equivalent to the 
computational basis and eigenbasis, which are ideal candidates for comparisons and 
benchmarking. Here we want to analyze the parameter space of the maple 
leaf lattice to investigate the performance of basis optimization in different 
regions. 

The maple-leaf lattice is a $1/7$-depleted triangular lattice. The deletion of 
sites leaves us with a triangular lattice of hexagons, surrounded by `belts' of 
triangles (see \figref{fig:1}{a}), with a coordination number of every
site of $z=5$, responsible for the name `maple-leaf'.
Whereas all sites are symmetry equivalent, the maple-leaf lattice naturally 
includes three types of symmetry-inequivalent nearest-neighbor bonds, offering an 
ideal playground to test our method since each type of bond in isolation 
decouples the lattice into products of dimers, triangles or hexagons (see 
\figref{fig:1}{a}).

We define the Hamiltonian on the maple-leaf lattice as
\begin{equation}
\mathcal{H} = J_d \sum_{\langle ij \rangle_d} \mathbf{S}_i\mathbf{S}_j + J_t \sum_{\langle ij \rangle_t} \mathbf{S}_i\mathbf{S}_j + J_h \sum_{\langle ij \rangle_h} \mathbf{S}_i\mathbf{S}_j
\end{equation}
where $J_d$, $J_t$, $J_h$ are the three symmetry-inequivalent bonds that run 
along the dimer, triangular, and hexagonal neighbors, respectively (see 
\figref{fig:1}{a}). We set the energy scale by normalizing $J_d^2 +
J_h^2 + J_t^2 = 1$ and parametrize the couplings by
\begin{equation}
\begin{aligned}
J_h &= \sin(\theta)\sin(\phi), \\
J_t &= \sin(\theta)\cos(\phi), \\
J_d &= \cos(\theta).
\end{aligned}
\end{equation}
where $\theta,\phi \in [0,\pi/2]$, since we only include antiferromagnetic 
couplings.  

This model has several known limits and has also been studied along different parts in the parameter space. As mentioned above, when only one of the couplings is different from zero the 
system decouples into isolated clusters: isolated dimers for $J_d\neq 0$, 
isolated trimers for $J_t \neq 0$, and isolated hexagons for $J_h\neq 0$. These limits are exactly solvable and
thus can be rendered sign-problem free for QMC within the computational basis 
(only for the hexagonal or dimer limits) or the cluster eigenbasis (for all 
three limiting cases). Therefore, we expect these bases to perform well in the 
proximity of these limits. In fact, the product state of singlets along the dimers is the exact ground 
state also for small but non-zero $J_t=J_h$~\cite{Ghosh2022}. Apart from these limiting 
points, the $J_t=0$ (i.e.\ $\phi = \pi/2$) line can be thought of as a honeycomb 
lattice with hard (all couplings identical) and mixed ($J_h$ and $J_d$ 
couplings) hexagons. As such, it is completely sign-problem free in the 
computational basis and has been studied in Ref.~\cite{Adhikary2021}, where a 
small Néel-ordered phase between the hexagonal singlet and dimer phases was 
identified. Another line that has been studied is the $J_h = 0$ ($\phi = 0$) 
limit, which corresponds to a distorted and frustrated star lattice. The 
thermodynamic properties were studied with QMC in Ref.~\cite{Reingruber2024} 
using the computational, dimer, and trimer bases. A very recent exact diagonalization study~\cite{Ebert2026} identified signatures of valence-bond-solid, Néel and $120^\circ$ Néel phases, 
dimer phases, and hexagonal plaquette phases. A projective symmetry group analysis~\cite{Sonnenschein2024} has further classified the possible U(1) and $\mathbb{Z}_2$ quantum spin liquids that could exist in this lattice.

\subsection{Sign problem in the cluster eigenbasis}

We first analyze and compare how different cluster eigenbases perform throughout the whole parameter space.  For this, we use the computational basis, the two-site eigenbasis defined on the dimers along the amber bonds (forming an effective kagome superlattice), and the three-site eigenbasis defined on the triangles along the teal bonds (forming a honeycomb superlattice). One could also consider the six-site eigenbasis running along the blue hexagons; however, since the corresponding supersite is a $64$-dimensional object, the SSE algorithm we employ becomes prohibitively slow.

\begin{figure}[t]
\centering
\includegraphics[width=\textwidth]{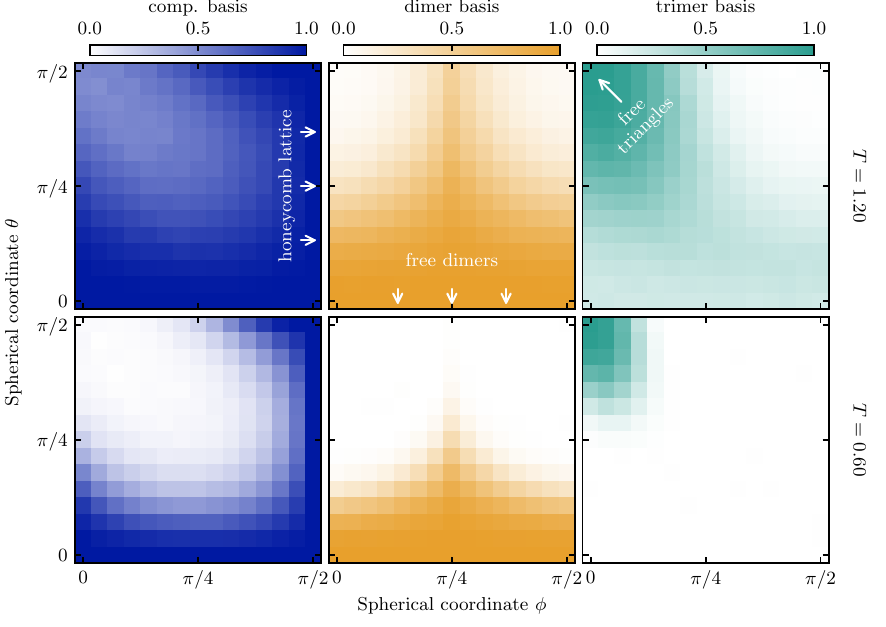}
\caption{Average QMC sign for the maple-leaf lattice on the $(\theta,\phi)$
grid, obtained from simulations in three different bases at $T = 1.2$ and
$0.6$. The lattice consists of $N = 54$ sites with periodic boundary conditions
(see \autoref{sec:appendix_full_cluster} for the full cluster).
The limiting cases of free dimers ($\theta = 0$), free trimers ($\theta = \pi/2, \phi = 0$), and the honeycomb lattice ($\phi = \pi/2$) are indicated in white.} 
\label{fig:sign_prob_bases}
\end{figure}

The average QMC sign $\langle s \rangle$ results are shown in 
\autoref{fig:sign_prob_bases}, with each basis on each different column: 
computational, two-site and three-site eigenbasis, respectively from left to 
right. Each row represents a different temperature, $T=1.2$ and $0.6$. 
The three bases exhibit a good sign performance in different regions of the parameter space, although lowering the temperature always leads to a shrinkage of high-sign regions. Particularly, the 
computational basis performs well in the $\theta=0$ limit, which corresponds to 
the sign-problem free dimer limit, and the $\phi=\pi/2$ limit where $J_t=0$ and 
there is no frustration. We clearly see that the performance of the computational basis
deteriorates strongly in the presence of the frustrating triangular interaction 
$J_t$, which becomes dominant in the vicinity of $\theta = \pi/2$ and $\phi = 
0$. On the contrary, the three-site eigenbasis (right column) has the best performance in regions where the system is dominated by $J_t$ interactions, and quickly fails 
outside this corner of the phase diagram. Finally, the dimer basis (middle column) yields larger average signs when $J_d$ dominates at $\theta=0$, but also extends to $\theta \neq 0$ especially along the line with $\phi = \pi/4$ (or $J_t = J_h$). This large-sign region in the dimer eigenbasis seems consistent with the large dimer phase.

\begin{figure}[t]
    \centering
    \includegraphics[width=1\textwidth]{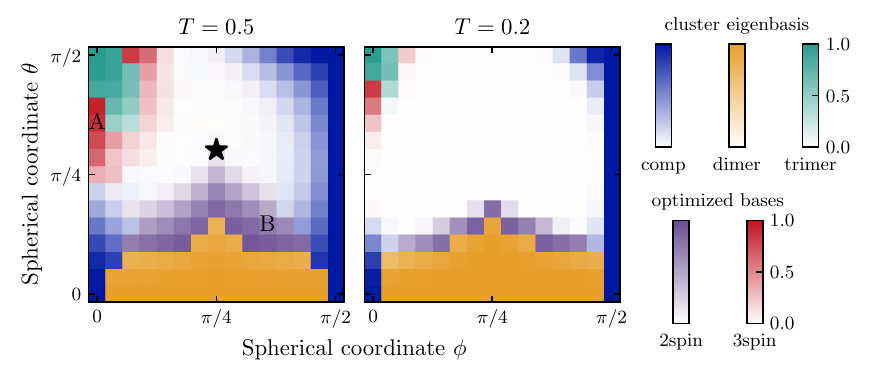}
    \caption{Computational phase diagram of the MLL: Average QMC sign of the
    computational/dimer/trimer basis is represented in blue/amber/teal,
whereas the average QMC sign of the optimized basis within the 2-spin/3-spin clustering is 
shown in purple/red. The plots show always the sign of the basis with the 
largest value of the sign in the respective parameter region. The simulation was performed on an $N = 54$ site lattice with periodic boundary conditions. The isotropic point $J_h=J_t=J_d$ is indicated by the black star. Left and right panel correspond to $T=0.5$ and $T=0.2$, respectively.}
    \label{fig:comp_phase}
\end{figure}

\subsection{Basis optimization and computational phase diagram}

\begin{figure}[ht]
\centering
\includegraphics[width=\textwidth]{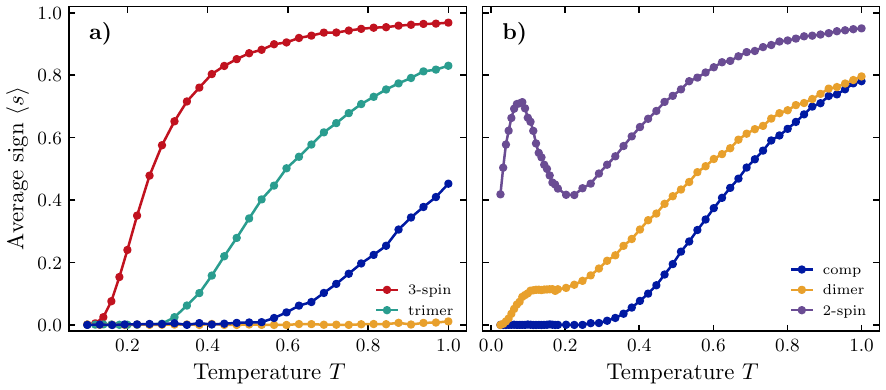}
\caption{Average QMC sign against temperature for the cluster eigenbasis 
(computational, dimer, and trimer bases) as well as the best bases we 
obtained from optimization (2- and 3-spin clustering). The simulation is carried 
out on an $N = 54$ MLL with $\theta = 11\pi/28$, $\phi = 0$ (Left) and 
$\theta = \pi/7$, $\phi = 5\pi/14$ (Right). These two parameter points are indicated in \autoref{fig:comp_phase} as A/B.}
\label{fig:star_line}
\end{figure}

We now apply the basis optimization method to the MLL. For this, we consider two clustering schemes: (i) two spins along the amber bonds and (ii) three spins along the teal bonds. For scheme (i), we optimize a single basis rotation $U \in \mathrm{U}(4)$. For scheme (ii), the three-spin clusters life on a bipartite effective superlattice, so we minimize the non-stoquasticity over $U_A \times U_B \in \mathrm{U}(8)\times \mathrm{U}(8)$. For every $(\theta,\phi)$ on the grid, we perform multiple optimizations with 100 random initializations and select the basis that yields the best sign performance.

To assess how well the optimization method performs, we introduce the \emph{computational phase 
diagram}. It maps, across the entire $\theta, \phi$ parameter grid, which of the 
five bases — computational, dimer, trimer, optimized two-spin, and optimized 
three-spin — achieves the highest average QMC sign $\langle s \rangle$ at each 
grid point, represented as a five-colored heat map (see \autoref{fig:comp_phase}). Interestingly, the method finds non-trivial bases that outperform both the cluster eigenbases and the computational basis, as indicated by the red and purple regions in \autoref{fig:comp_phase}. These bases appear near the boundaries of the dimer- and trimer-dominated regions. On the other hand, close to the limits of isolated dimers or trimers, our method finds that the corresponding cluster eigenbases remain optimal. Overall, in most of the phase diagram the average sign still decreases towards zero as the temperature is lowered (as can be seen in the right panel at $T=0.2$).

Along the star-lattice line ($\phi = 0$), the improvement achieved by the 
optimization is most pronounced. As shown in \figref{fig:star_line}{a}, the best basis within the 3-spin clustering enables 
simulations to reach significantly lower temperatures for a fixed system size 
(in our case $N = 54$). While the optimized basis allows us to reach $T = 0.15$, 
the trimer basis cannot go below $T = 0.3$, and the computational basis merely 
down to $T \approx 0.55$. This constitutes a clear improvement over the 
simulation results for the Heisenberg antiferromagnet on the star lattice presented in 
Ref.~\cite{Reingruber2024}, where only cluster eigenbases were employed. 

In \figref{fig:star_line}{b}, we show that our method also makes considerable improvements in the dimer phase. Specifically, we show this for $\phi = 5\pi/14, \theta = \pi/7$. It is striking that at this parameter point the dimer basis simulation yields a plateau in the average sign at $T \approx 0.2$ (amber curve), whereas the optimized basis shows there a inflection point (purple curve). 

\subsection{Sign problem in the dimer phase}

\label{sec:dimer_bas_dimer_phase}

\begin{figure}[t]
\centering
\includegraphics[width=1\textwidth]{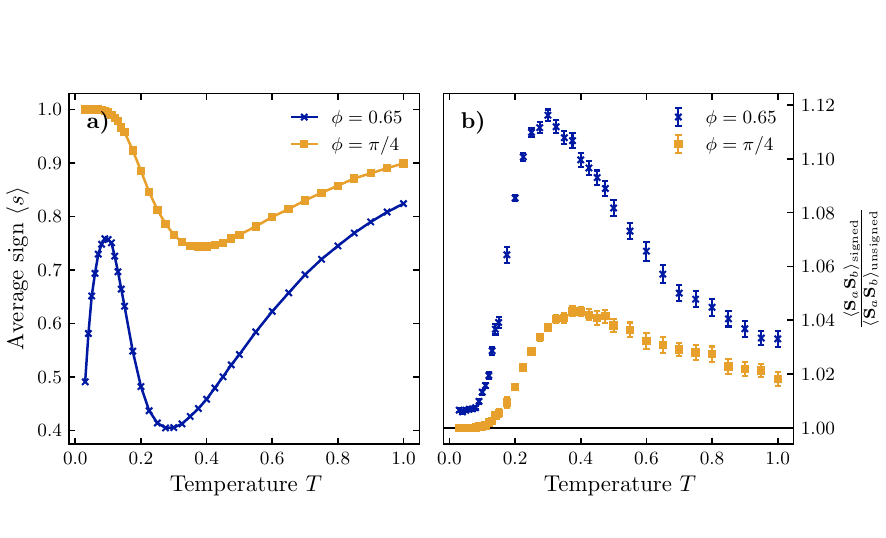}
\caption{(Left) QMC sign against temperature on the MLL at $\theta = 0.5$ simulated in the dimer eigenbasis for two values of $\phi$ on an $N=54$ lattice. (Right) Ratio of the dimer correlation $\langle S_a\cdot S_b\rangle$ obtained from the signed and unsigned Hamiltonians for the same parameters.}
\label{fig:sign_recovery}
\end{figure}

We observe that in some cases the average sign increases at 
low temperatures. A similar behavior was reported in 
Refs.~\cite{Honecker2022,Wessel2018} for the Shastry--Sutherland model in the 
weakly-coupled dimers limit, where the ground state is an exact product state of 
singlet dimers. In these works, it was observed that the average sign recovers 
completely to $1$ at zero temperature, allowing for sign-problem free 
simulations. We observe this behavior for $\phi=\pi/4$, corresponding to $J_h = 
J_t$, as can be seen in \figref{fig:sign_recovery}{a} (amber
curve). In Refs.~\cite{Honecker2022,Wessel2018}, the recovery was associated to 
the ground states of the signed and unsigned Hamiltonians being the same (see 
\autoref{eq:average_sign_formula}), thus producing that at low temperatures 
$\Delta f \simeq \Delta e = 0$.

Interestingly, for $\phi \neq \pi/4$ ($J_h\neq J_t$), we observe a different behavior: $\langle 
s\rangle$ initially recovers upon lowering the temperature but vanishes again at 
very low temperatures (see blue curve in \figref{fig:sign_recovery}{a} and purple curve in \autoref{fig:star_line}). 
Namely, it has a local minimum and maximum at low temperatures. To the best of 
our knowledge, such behavior has not been reported in the literature so far. It 
can be understood from the following considerations: We already mentioned that 
the sign recovers when the unsigned Hamiltonian and the physical Hamiltonian 
share the same ground state~\cite{Honecker2022,Wessel2018,Pan2024}. In the 
dimerized phase ($J_h, J_t \ll J_d$), the ground state of the physical Hamiltonian is dominated by 
the product of singlets along the amber dimer bonds, which is an exact
eigenstate of the Hamiltonian along the $J_t = J_h$ line (also for the unsigned 
Hamiltonian in the dimer eigenbasis). The 
eigenstate property can be made plausible by looking at the interaction between 
two dimers $d$ and $d^\prime$, which is $h_{dd^\prime} = J_t\, \mathbf{S}_a 
\mathbf{S}_b^\prime + J_h\, \mathbf{S}_b \mathbf{S}_b^\prime$ (see 
\figref{fig:1}{a}), being $a$ and $b$ the two spins in a given dimer. We
can then re-write the interaction in terms of the total spin 
$\mathbf{T}_d^{(\prime)} = \mathbf{S}_a^{(\prime)} + \mathbf{S}_b^{(\prime)}$ 
and the spin difference $\mathbf{D}_d^{(\prime)} = \mathbf{S}_a^{(\prime)} - 
\mathbf{S}_b^{(\prime)}$, resulting in
\begin{equation}
h_{dd^\prime} = \frac{J_t + J_h}{4}\left(\mathbf{T}_d \mathbf{T}_d^\prime - \mathbf{T}_d \mathbf{D}_d^\prime \right)
   + \frac{J_t - J_h}{4}\left(\mathbf{D}_d \mathbf{T}_d^\prime - \mathbf{D}_d \mathbf{D}_d^\prime \right).
\label{eq:dimdim}
\end{equation}
Now, the singlet is an eigenvector of $\mathbf{T}_d$ with eigenvalue 0 and it is not an eigenvector of $\mathbf{D}_d$, which turns it into a triplet. Starting from the trivial pure singlet state at $J_t=J_h=0$, adding a small $J_t=J_h$ does not change the state because the second term in \autoref{eq:dimdim} vanishes due to the coupling constants being equal and the first term vanishes due to the $\mathbf{T}_d$ operators acting on the singlets. If we deviate from this line and make $J_t \neq J_h$, the second term has to be taken into account, and the $\mathbf{D}_d \mathbf{D}_d^\prime$ part promotes the existence of triplets, destroying the pure singlet product state. Thus, it is plausible that there is no complete sign recovery for $\phi \neq \pi/4$.

To understand why the sign exhibits a maximum for $\phi \neq \pi/4$, we show in \figref{fig:sign_recovery}{b} the ratio of the spin-spin correlations in the physical and unsigned Hamiltonians,
\begin{equation}
\frac{\langle S_a\cdot S_b\rangle_\text{signed}}{\langle S_a\cdot S_b\rangle_\text{unsigned}}
\end{equation}
as a function of temperature for both $\phi = 0.65$ and $\phi = \pi/4$. The 
ratio approaches exactly $1$ for $\phi = \pi/4$ at low temperatures, indicating 
that both Hamiltonians share the same ground state. On the other hand, for $\phi 
\neq \pi/4$ the ratio does not reach $1$, signaling that even though both ground 
states remain strongly dominated by dimer singlets $\left(\langle S_a\cdot S_b 
\rangle \approx -0.75\right)$, they are different ($\Delta e \neq 0$). Because of this 
small but finite difference, the average sign eventually vanishes for $T \to 0$ 
as the SSE expansion order grows. To conclude, this occurs because the ground 
state is no longer an exact product state of singlets for $J_h\neq J_t$.

The local minimum and maximum of the sign in \autoref{fig:sign_recovery} can be attributed to the interplay of two competing effects. Upon cooling, the state ensembles of the physical and unsigned Hamiltonians become progressively more different. This is reflected by the behavior of the correlation ratio, which reaches a maximum deviation from $1$ around \(T \approx 0.25\). This is where the minimum of the average sign appears. When lowering the temperature, the ensembles become more and more similar again as the ratio approaches $1$. Thus, $\Delta f \ll 1$ and fewer negative signs accumulate in the operator strings, leading to a partial recovery of the average sign, \(\langle s \rangle = Z/Z^\prime\). Finally, lowering the temperature further causes the sign to vanish because $\Delta f \neq 0$ and $\beta \to \infty$ in \autoref{eq:average_sign_formula}.

\subsection{Comparison to NLCE}

\begin{figure}[t]
    \centering
    \includegraphics[width=1\textwidth]{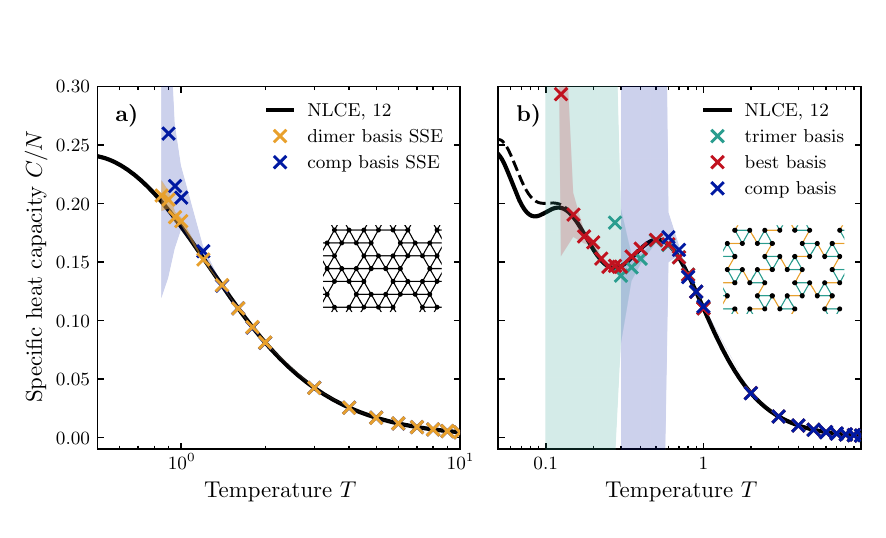}
	\caption{Comparison of NLCE and QMC data for the specific heat capacity. The
    NLCE data obtained from a triangle-based expansion (order 12 solid line, order 11 dashed line) and is shown in black. Details can be found in Ref.~\cite{Schafer2026}. The QMC was carried out on an $N = 54$ site lattice. Errorbars are shown as shaded regions to avoid clutter. (Left) comparison of the two methods at the isotropic point. (Right) at $\theta = 1.122$ and $\phi = 0$.}
    \label{fig:heatcap_maple_compa}
\end{figure}

We benchmark our QMC data against NLCE, an unbiased and controlled finite-temperature method in \autoref{fig:heatcap_maple_compa}. NLCE computes thermodynamic observables directly in the thermodynamic limit by systematically including the contributions of finite subclusters up to a given size (order). We refer the reader to Refs.~\cite{Tang2013, Schafer2020, schaefer_magnetic_2022} for further details.
The method was applied to the MLL in Ref.~\cite{Schafer2026}, where the specific heat was computed at the isotropic point $J_h = J_t = J_d$, finding a two-peak structure signaling the presence of two distinct energy scales in the system. They argue for the possibility of a paramagnetic hexagonal plaquette ground state.

\paragraph{Isotropic point.}
At $J_h=J_d=J_t$, our optimization method identifies the dimer eigenbasis as the most efficient
choice with respect to the QMC sign. As shown in \figref{fig:heatcap_maple_compa}{a}, for a fixed computation time, simulations performed in the dimer eigenbasis (amber crosses) yield noticeably smaller error bars than those performed in the computational basis (blue crosses). Even though the cluster eigenbasis has already been used on other systems before~\cite{Wessel2018,Reingruber2024,Weber2022a,DEmidio2020}, in the present case none of the three 
symmetry-inequivalent bonds is preferred, and it is therefore not evident 
\emph{a priori} that the dimer basis should outperform the computational basis. 
This is where the optimization approach becomes particularly valuable, as it 
provides a systematic way to identify an optimal basis without requiring prior
knowledge of the underlying physics.

\paragraph{Star-lattice point $(\theta = 1.122,\ \phi = 0)$.}
At the parameter point $(\theta = 1.122, \phi = 0)$ corresponding to $(J_h, J_t, J_d) \approx (0, 0.901, 0.433)$, our method finds an optimized basis that is neither the computational nor the cluster eigenbasis. The resulting data for the specific heat capacity are shown in \figref{fig:heatcap_maple_compa}{b} and compared with the NLCE calculations. With the same computational effort, the optimized basis (red crosses) clearly outperforms the trimer basis (teal crosses), which in turn outperforms the computational basis (shown in blue). The specific heat exhibits a broad minimum in the range $T \in [0.15, 0.3]$, which can only be resolved using the optimized basis. This demonstrates a physically relevant improvement of the simulation. These results are an independent confirmation of the Euler-resummed NLCE results which are converged in the same temperature regime that we can reach with sign-optimized QMC. The agreement of thermodynamic limit NLCE and $N=54$ QMC results demonstrates furthermore that the minimum of the specific heat capacity is a robust feature 
likely generated by short-range correlations which are already well represented by the $N=54$ cluster.
For NLCE, convergence is indicated by consistency between successive orders, i.e., where the dashed and solid lines of the same color overlap.

In \autoref{ch:more_benchmark} we present the comparison of the 
two methods for three more parameter points. While finite order NLCE is a 
physically motivated high-temperature expansion, the application of series 
acceleration methods to the NLCE series defies physical intuition. Our benchmark 
with a completely independent and state-of-the-art QMC method enhanced by our 
sign optimization demonstrates that this series extrapolation is indeed able to 
access length scales beyond the size of the finite-size clusters included in 
NLCE and competitive with QMC calculations on system sizes beyond reach for 
exact diagonalization. These results can help to gauge expectations at other 
parameter points. It is interesting that at parameter point $(\theta = 1.122, \phi = 0)$, the converged temperature range of NLCE 
is comparable to the regime accessible with best-effort sign-optimized QMC 
calculations and that neither method has a large advantage. 

Overall, we demonstrate that the basis-optimization approach provides 
a systematic way to improve the quality of QMC. This is necessary to obtain 
results competitive with NLCE. Naive computational basis QMC is easily 
outperformed by high temperature series expansions in this challenging MLL 
system. The enhanced accessible temperature range and reduced statistical errors 
allow the basis-optimized QMC to serve as a state-of-the-art benchmark. Since 
the optimization increases the sampling efficiency, finite-temperature regimes 
that are inaccessible by zero-temperature methods can now be studied with 
reasonable statistical accuracy at an affordable computational cost.

\section{Conclusion}
\label{sec:conclusion}

Our investigation shows that the sign problem can in many cases be alleviated by 
a suitable local unitary basis rotation, even for challenging two-dimensional 
frustrated magnets. While it is generally unrealistic to expect that a local 
basis will allow sign free simulations, it is nevertheless encouraging that 
accessible temperatures can be pushed down by about a factor four (compared to the computational basis) in some cases, 
resolving nontrivial features otherwise inaccessible.

We extended the closely related approach of Ref. \cite{Hangleiter2020} to 
leverage the entire unitary group for the basis rotation, by generalizing the 
cost function for complex-valued Hamiltonians. This is required for representing 
the well known Marshall-sign transformation, which our approach recovers.
In the cases where our optimal bases differ from the eigenbasis of local 
building blocks used in previous pioneering work \cite{Wessel2017,Weber2022a}, we observe 
that the expressive power of unitaries over orthogonal rotations yields an 
advantage.

Our analysis revealed that this cost function is closely tied to the average 
sign, and that the extrema of both quantities appear to coincide in all cases we 
considered. This is somewhat surprising and very useful since the sign is a 
noisy QMC observable while the stoquasticity can be determined by inspection of 
the rotated Hamiltonian. Furthermore, we find that (generalized) gradient 
descent optimization is easier for our cost function compared to the sign both 
due to the noise and rugged landscape of the latter. The fact that the extrema 
coincide is also reflected in the alignment of the gradients of both quantities 
during optimization close to the extrema.

For the fully-frustrated ladder, the optimization consistently recovers a stoquastic basis,
reproducing the known exact sign-free representation at $\mathcal{S} = 0$. For the
triangular ladder, we were unable to find a sign-problem-free basis within the two-site cluster ansatz,
but we identified unitary rotations that significantly improve the sign compared to both
the cluster eigenbasis and the computational $S^z$ basis.

For the maple-leaf lattice, we mapped out a computational phase diagram over the full
two-dimensional coupling-anisotropy parameter space, identifying regions where bases resulting from optimization achieve an average sign larger than the cluster eigenbases or the $S^z$ eigenbasis. The improvements emerge near the boundaries between dimer- and trimer-dominated regions. The most pronounced improvement appears along the star-lattice line, where
the optimized three-site basis extends the accessible temperature range from
$T \approx 0.3$ to $T \approx 0.15$ compared to the trimer eigenbasis, enabling
improved benchmarking against state-of-the-art NLCE and the resolution of a nontrivial low-temperature minimum in the specific heat.

The improvement is not uniform across parameter space. Near the isotropic point, where
the sign problem is believed to be most severe, local basis rotations become less effective, as the
underlying physics is highly non-local. These regimes are unfortunately often of greatest
physical interest due to the strong competition between interactions. We do not 
expect the method to substantially extend the accessible
temperature range for large system sizes in this regime.

At high temperatures, we find convincing improvement over a large part of the 
parameter space and we expect this to be true quite generically. At lower 
temperatures, we speculate that locally optimized bases yield significant 
improvements over the computational basis if the low energy part of the full 
eigenbasis is weakly entangled. This follows from the observation that for 
perfect (dimer-/triangle- etc.) product states, the sign can be cured. A natural 
extension of this fact seems to be by low virtual dimension tensor networks 
interpolating between product states and strongly entangled wave functions.
Future work could therefore investigate this aspect in situations where 
couplings between larger building blocks are weak, such as in breathing 
lattices, e.g. the breathing kagome or pyrochlore lattice, where the 
optimization could be performed on the strong triangles \cite{machado2026} or 
tetrahedra~\cite{Harris1991,Schafer2021}.

Generally, systematic basis optimization provides a robust tool to improve 
sampling efficiency, and we recommend combining
different clustering schemes with multiple random-initialization optimization runs to
identify the most efficient basis within a given cluster ansatz.

It would also be worthwhile to explore alternative cost
functions beyond non-stoquasticity and to investigate whether the optimized bases encode
physical information about the system.

\section*{Acknowledgements}
We thank Paul L. Ebert and Alexander Wietek for providing data from their studies of the phase diagram of the maple-leaf antiferromagnet.

\paragraph{Funding information}
We acknowledge support by the DFG through the cluster of excellence
ML4Q (EXC 2004, project-id 390534769), CRC1639 NuMeriQS (project-id 511713970) 
and CRC TR185 OSCAR
(project-id 277625399).
R. S. acknowledges support from the DFG under Project No. 575641691, the Helmholtz-Zentrum Berlin and the Freie Universität Berlin.

\begin{appendix}
\numberwithin{equation}{section}

\section{Parametrization of unitary matrices}
\label{sec:appendix_givens}

In this appendix we provide the explicit parametrization of unitary matrices 
$U \in \mathrm{U}(4)$ used in the optimization procedure, based on the general 
principle of factorizing a unitary matrix into a product of two-dimensional 
rotations and diagonal phase matrices, as in Refs.~\cite{Clements2017,Reck1994}.

\subsection*{General construction}

We represent a general unitary matrix $U \in \mathrm{U}(4)$ as a product of elementary
two-dimensional unitary rotations (complex Givens rotations), supplemented by
a diagonal phase matrix. This construction guarantees unitarity by construction
while allowing for an unconstrained parametrization in terms of real variables.

The parametrization depends on a real parameter vector
\begin{equation}
\theta = (\{\theta_k\}_{k=1}^6, \{\phi_k\}_{k=1}^6, \{\delta_i\}_{i=1}^4) \in \mathbb{R}^{16}.
\end{equation}

The unitary is constructed as
\begin{equation}
U(\theta) = \left( \prod_{k=1}^{6} G_{p_k q_k}(\theta_k, \phi_k) \right) D_{\mathrm{in}}(\{\delta_i\}),
\end{equation}
where the product is ordered from right to left. The diagonal phase matrix is given by
\begin{equation}
D_{\mathrm{in}} = \mathrm{diag}\left(e^{i\delta_1}, e^{i\delta_2}, e^{i\delta_3}, e^{i\delta_4}\right).
\end{equation}
Each elementary rotation $G_{pq}(\theta,\phi) \in \mathrm{U}(4)$ acts non-trivially only
on the two-dimensional subspace spanned by basis states $p$ and $q$.
It is defined as
\begin{equation}
G_{pq}(\theta,\phi) =
\begin{pmatrix}
\ddots &        &        &        \\
       & e^{i\phi}\cos\theta & e^{i\phi}\sin\theta &        \\
       & -e^{-i\phi}\sin\theta & e^{-i\phi}\cos\theta &      \\
       &        &        & \ddots
\end{pmatrix},
\end{equation}
where the $2\times2$ block acts on indices $(p,q)$ and all other matrix elements
coincide with the identity. Explicitly, the entries are
\begin{align}
(G_{pq})_{pp} &= e^{i\phi}\cos\theta, \\
(G_{pq})_{pq} &= e^{i\phi}\sin\theta, \\
(G_{pq})_{qp} &= -e^{-i\phi}\sin\theta, \\
(G_{pq})_{qq} &= e^{-i\phi}\cos\theta.
\end{align}
This parametrization ensures $G_{pq} \in \mathrm{U}(4)$ for all real $\theta$ and $\phi$.

\subsection*{Choice of rotation sequence}

We employ a fixed sequence of index pairs $(p_k, q_k)$ corresponding to a column-wise (upper-triangular) Givens decomposition:
\begin{equation}
(p_k, q_k) \in \{(1,2), (1,3), (1,4), (2,3), (2,4), (3,4)\}.
\label{eq:A9}
\end{equation}
The full unitary is therefore constructed as
\begin{equation}
U = G_{34} \, G_{24} \, G_{23} \, G_{14} \, G_{13} \, G_{12} \, D_{\mathrm{in}}.
\end{equation}

\subsection*{Remarks}

The chosen ordering of Givens rotations is not unique; different orderings
correspond to different coordinate systems on $\mathrm{U}(4)$.

Each parameter affects only a two-dimensional subspace, leading to a localized parametrization. In contrast, the exponential map $U = e^{iH}$ with a Hermitian generator $H$ introduces highly nonlocal 
dependencies between parameters and matrix elements. An additional advantage of the Givens-rotation parametrization concerns the 
interpretability of gradients, which is particularly useful for the analysis of gradients in \autoref{sec:grads}. Variations with respect to individual parameters generate well-defined and localized directions on $U(4)$. In contrast, for an exponential parametrization $U = e^{iH}$, the geometric meaning of individual parameter directions remains unclear. 

\section{Histogram of Optimization Results}
\label{sec:appendix_histogram}
\begin{figure}[ht]
    \centering
    \includegraphics[width=0.7\textwidth]{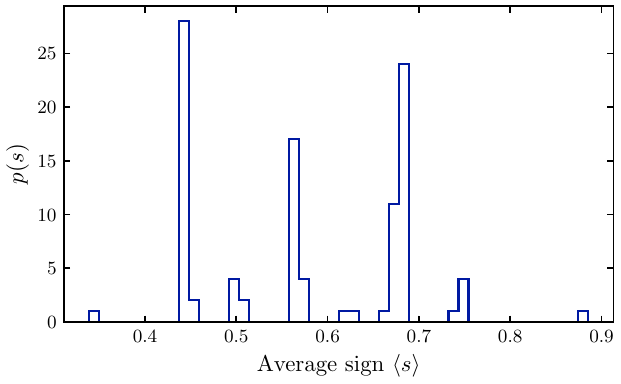}
    \caption{Histogram of the average QMC sign values at temperature $T=1.2 J$ for the optimized bases. The eight distinct classes of bases with respect to $\langle s \rangle$ are represented by the eight peaks. The computational basis is represented by the large peak at $\langle s \rangle \approx 0.45$. The simulations were carried out on a triangular ladder with $N=24$ sites.} 
    \label{fig:hist}
\end{figure}

In order to illustrate that the bases resulting from the optimization can be classified into distinct families with respect to their sign performance, we show in \autoref{fig:hist} a histogram of the average QMC sign values at temperature $T=1.2 J$ for the optimized bases.
\clearpage
\section{Landscape}
\label{ch:landscape}

\begin{figure}[t]
    \centering
    \includegraphics[width=0.7\textwidth]{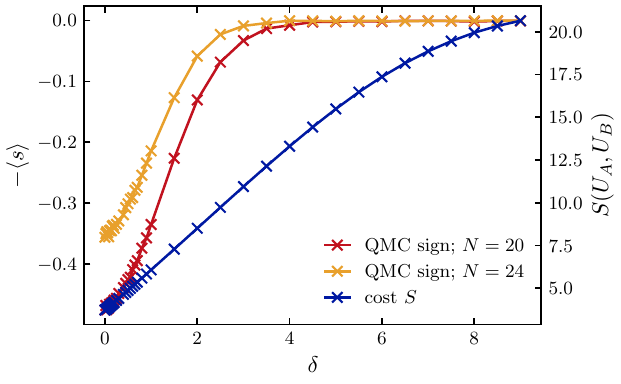}
	\caption{Non-stoquasticity $\mathcal{S}$ and negative value of the average sign $-\langle s \rangle$ along a straight line through the parameter space starting from the cost minimum, parametrized by $\delta \in \mathbb{R}$. The sign is calculated for system sizes $N=20$ and $N=24$. The simulations are carried out at $T=J$ on a triangular ladder.}
    \label{fig:sign_cost_cross_section}
\end{figure}

In \autoref{fig:sign_cost_cross_section}, we show the average sign and the 
evolution of the cost function starting from the optimal value along a line in 
parameter space parametrized by a parameter $\delta$. The blue curve corresponds 
to the non-stoquasticity $\mathcal{S}$ and the red and amber curves
correspond to the average sign $\langle s\rangle$ for two different system 
sizes. This result indicates that the optimization landscape of the cost 
function is better conditioned than that of the sign, providing evidence for the 
statement in the main text that the energy landscape of the cost function is 
better suited for optimization than the sign due to less sharp extrema.

\section{Benchmarking with NLCE at Additional Parameter Points}
\label{ch:more_benchmark}

\begin{figure}[ht]
    \centering
    \includegraphics[width=1\textwidth]{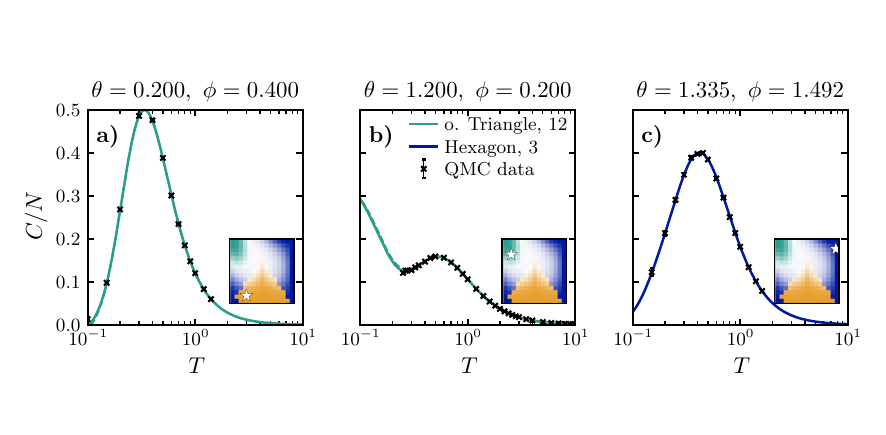}
	\caption{Comparison of NLCE (see Ref.~\cite{Schafer2026}) and QMC data for the specific heat capacity at three additional parameter points. The simulation was carried out on an $N = 54$ lattice in the respective best basis. The previous order in a certain expansion scheme is represented by a dotted line in the respective color.}
    \label{fig:more_benchmark_with_NLCE}
\end{figure}

Here we show additional results, where we compare the SSE-QMC data in the optimized basis with NLCE. The results are shown in \autoref{fig:more_benchmark_with_NLCE}, where we find that both numerical methods agree within uncertainties at all three parameter points. The NLCE algorithm used an expansion based on overlapping triangles and hexagon developed in Ref.~\cite{Schafer2026}.

\section{Full MLL cluster}
\label{sec:appendix_full_cluster}
\begin{figure}[ht]
    \centering
    \includegraphics[width=0.4\textwidth]{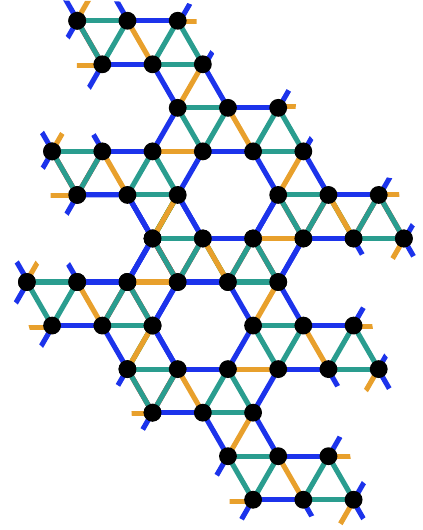}
    \caption{The full $N = 54$ site maple-leaf cluster used in our simulations, with periodic boundary conditions.}
    \label{fig:full_cluster}
\end{figure}

For all simulations on the maple-leaf lattice we use the $N = 54$ site cluster shown in \autoref{fig:full_cluster}, with periodic boundary conditions.

\end{appendix}

\clearpage
\bibliography{references}

\end{document}